# Applicability of scaling laws to vision encoding models

Takuya Matsuyama[1], Kota S Sasaki[1], Shinji Nishimoto[1, 2]

[1]Graduate School of Frontier Biosciences, Osaka University, Japan

[2]CiNet, NICT, Japan

## Abstract

In this paper, we investigated how to build a high-performance vision encoding model to predict brain activity as part of our participation in the Algonauts Project 2023 Challenge. The challenge provided brain activity recorded by functional MRI (fMRI) while participants viewed images. Several vision models with parameter sizes ranging from 86M to 4.3B were used to build predictive models. To build highly accurate models, we focused our analysis on two main aspects: (1) How does the sample size of the fMRI training set change the prediction accuracy? (2) How does the prediction accuracy across the visual cortex vary with the parameter size of the vision models? The results show that as the sample size used during training increases, the prediction accuracy improves according to the scaling law. Similarly, we found that as the parameter size of the vision models increases, the prediction accuracy improves according to the scaling law. These results suggest that increasing the sample size of the fMRI training set and the parameter size of visual models may contribute to more accurate visual models of the brain and lead to a better understanding of visual neuroscience.

## 1. Introduction

An intriguing topic in systems neuroscience is to build quantitative models that can describe the relationship between sensory experiences and brain activity. In particular, the construction of encoding models that predict brain activity from sensory features (e.g., visual shape and color) has been pursued with the goal of investigating visual processing in the brain. In recent years, deep neural networks (DNNs) have been used for such feature extraction (Güçlü & van Gerven, 2015; Wen et al., 2018). While DNNs enable the construction of encoding models with high predictive power, their construction typically requires a significant amount of data in two ways: first, to train the DNNs themselves, and second, to train the encoding models.

With the help of high-performance computing, advances in data science allow the latest DNN models to be built with an enormous number of parameters using gigantic datasets (e.g. Large Language Models). These models show exceptional performance, allowing them to be used in a wide range of applications. Their behavior is consistent with scaling laws, which state that model accuracy improves linearly with logarithm of the dataset size and parameter size (Henighan et al., 2020; Kaplan et al., 2020).

Does a similar scaling law apply to building encoding models of brain activity? When building encoding models for fMRI signals evoked by natural language stimuli, Antonello et al. (2023) have shown that prediction accuracy increases linearly on a logarithmic scale as a function of both the size of the samples used to train the encoding models and the size of the parameters included in the language models.

The above issue can be addressed in the visual domain today, as current large-scale vision models are evolving significantly to solve visual tasks, sometimes using non-visual cues (e.g., text, audio). Based on the latent representations inherent in modern vision models, we built encoding models for visual brain activity provided in the large-scale Natural Scene Dataset (NSD) by Allen et al. (2022). Specifically, this investigation asked quantitatively how the prediction accuracy of different encoding models depends on (1) the sample size used to construct the models and (2) the parameter size of the vision models.

## 2. Methods
### 2.1. Vision models
To construct encoding models to explain brain activity in visual cortices, we extracted latent representations from 6 vision models (Table 1). These models were selected because they have different architectures and were trained with different modalities (Table 1).

Table 1: Summary of the models used in this study

| Model name | Size name | Number of parameters | Architecture | Training modality |
|---|---|---|---|---|
| EVA-01-CLIP (Sun et al., 2023) | giant | 1.0B | Transformer | Image and Text |
| EVA-02-CLIP (Sun et al., 2023) | base | 0.086B | Transformer | Image and Text |
|  | large | 0.3B |  |  |
|  | enormous | 4.4B |  |  |
| ConvNext (Liu et al., 2022) | xxlarge | 0.85B | CNN | Image and Text |
| ONE-PEACE (P. Wang et al., 2023) | N / A | 1.5B | Transformer | Image, Text and Audio |

**2.2. fMRI data**

The dataset published by Allen et al. (2022) contains brain activity recorded by fMRI while participants viewed approximately 9000 to 10,000 images. This dataset was divided into training and test samples for each participant, as shown in Table 2. In this challenge, the test data provided only images without brain activity.

Table 2: The number of samples in the dataset provided by the challenge

|  | Subject01 | Subject02 | Subject03 | Subject04 | Subject05 | Subject06 | Subject07 | Subject08 |
|---|---|---|---|---|---|---|---|---|
| **Training** | 9841 | 9841 | 9082 | 8779 | 9841 | 9082 | 9841 | 8779 |
| **Test** | 159 | 159 | 293 | 395 | 159 | 293 | 159 | 395 |

**2.3. Encoding model construction**

We built voxel-wise encoding models to predict image-evoked brain activity (Naselaris et al., 2011). Brain activity **R** was modeled as

$$\mathbf{R} = \mathbf{SW} + \mathbf{e}$$

Where **S** are visual features extracted from vision models, **W** is a weight matrix, and **e** is noise. An L2-regularized linear regression (ridge regression) was used to optimize **W**. The regularization parameters were determined using the cross-validation method.

**2.4. Final model**

Prediction accuracy is generally improved by pooling the features in the intermediate layers of the vision models with an optimal kernel size (Li et al., 2022). In this study, max pooling was applied to the features extracted from each vision model.

We first determined the optimal combination of layer and kernel size for max pooling in each voxel using the training data through a cross-validation procedure. Then, using the combination of parameters that produced the most accurate prediction, we retrained the model using all training samples to predict brain activity in the test samples (see Table 2). The final predictions were generated by averaging the predictions of each vision model.

# 3. Results
## 3.1 Data size scaling

We examined how the prediction accuracy of our encoding models depended on the fMRI sample size used to build them. To equalize the number of samples across subjects, 8779 samples, the smallest of all subjects' training data, were used in the analysis. Using the features extracted from each layer of EVA-02-clip-large, we built different encoding models for each of the seven ROIs (anatomical streams). Figure 1 shows the prediction accuracy for each ROI, averaged across subjects. Prediction accuracy improved approximately log-linearly with increasing training samples for all subjects (Figure 1), suggesting that vision encoding models follow a scaling law with respect to the sample size of the fMRI training set.

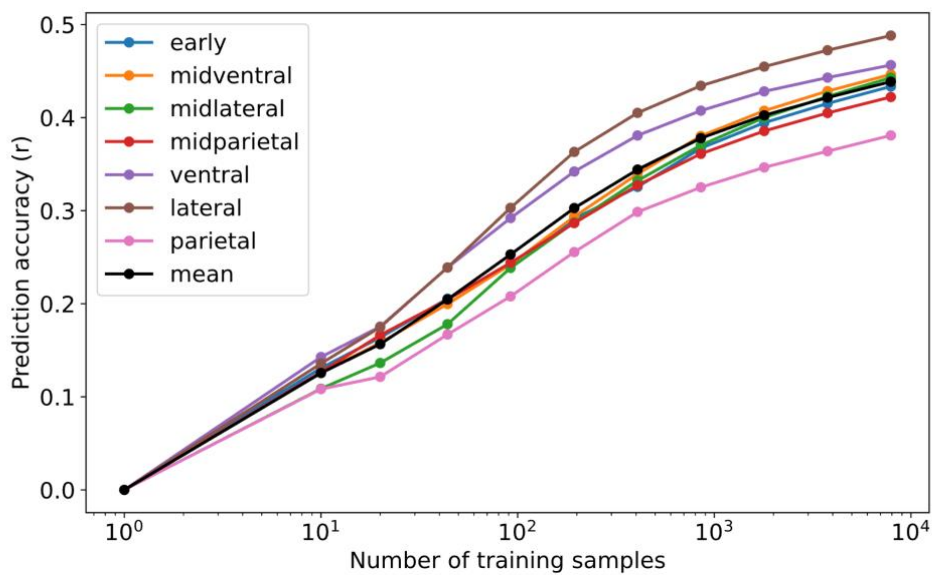

**Fig. 1 Scaling law for the sample size of the fMRI training set.**

Prediction accuracy (mean of correlation coefficients) of encoding models for validation data is shown for seven ROIs (color). The prediction accuracy averaged over these ROIs is also plotted as a black curve.

**3.2. Parameter size scaling**

We investigated how the prediction accuracy depends on the parameter size of the vision models while keeping the number of samples constant (4500 training samples and 500 validation samples). Figure 2 shows that the prediction accuracy for each ROI was averaged across subjects. When comparing all models, the prediction accuracy for each ROI does not appear to scale log-linearly. However, when we focus only on the EVA-02-CLIP series models, the prediction accuracy appears to scale log-linearly, especially in the higher order visual fields such as ventral, lateral, and parietal. This result suggests that a scaling law is likely applied to the parameter size of the vision model used in the vision encoding models. Comparison of all vision models suggests that in addition to scaling parameter sizes, attention to factors such as architecture, training methods, and training data quality can further improve the prediction accuracy of the encoding model.

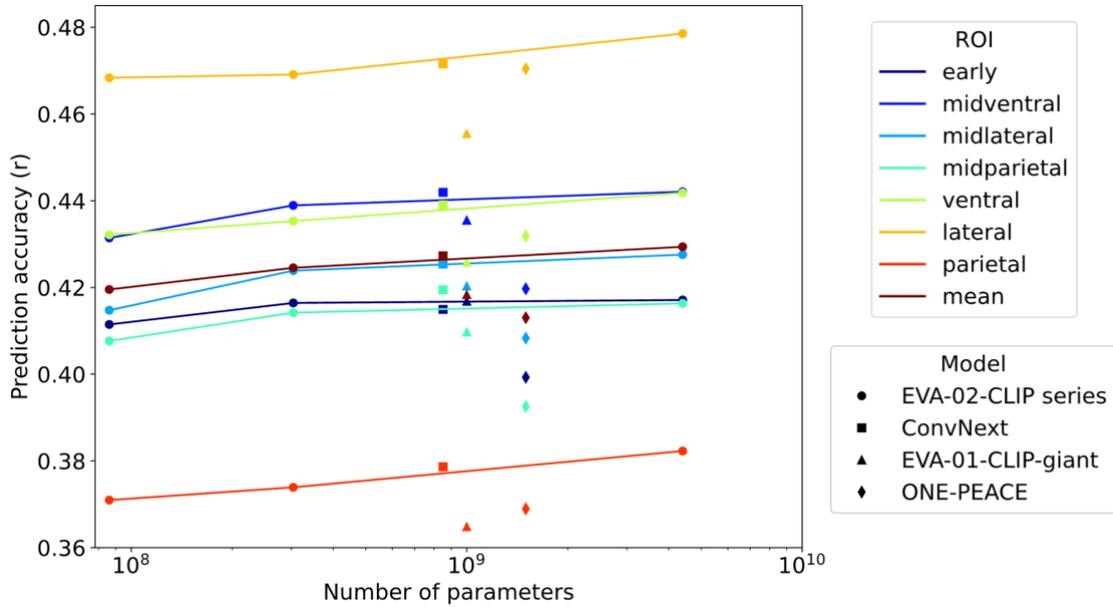

**Fig. 2 Scaling law for the parameter size of the vision models**

Prediction accuracy (mean correlation coefficient) for validation data is plotted for different vision models. The colors represent the stream ROI and the icons represent the vision model (EVA-02-CLIP series includes EVA-02-CLIP-base, EVA-02-CLIP-large, and EVA-02-CLIP-enormous). Plots are connected by line segments only for EVA-02-CLIP series.

### 3.3. Final submission results

Figure 3 shows the final prediction results for the Algonauts Project 2023 (Gifford et al., 2023). Our model predicted not only the early retinotopic visual regions (prf-visualrois), but also the higher visual regions, such as body selective regions (floc-bodies), face selective regions (floc-faces), word selective regions (floc-words), and place selective regions (floc-places) with about 60% accuracy. These results indicate that visual areas in all hierarchical ranks were predicted to a similar extent by encoding models constructed using large-scale parameters in vision models and large-scale fMRI samples.

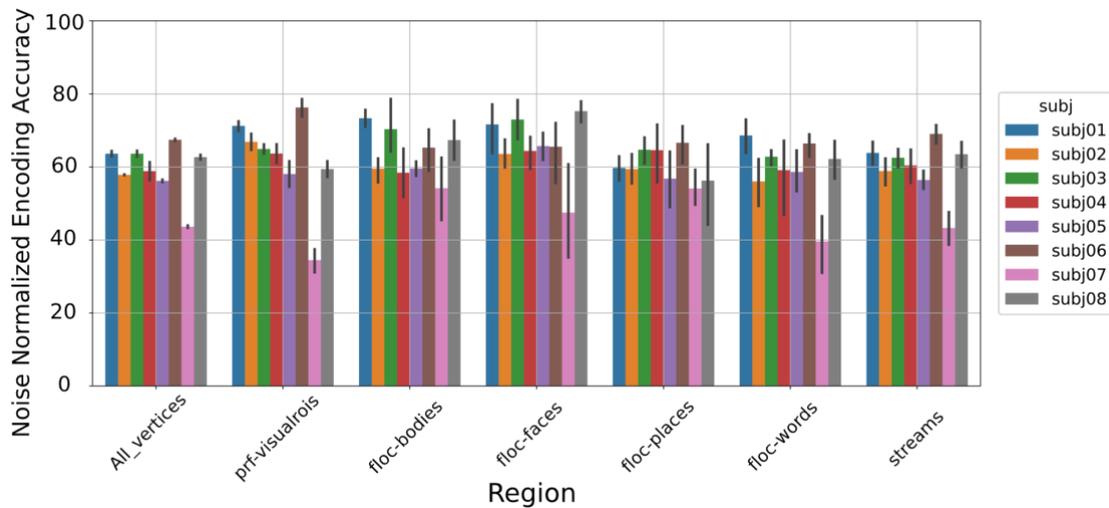

**Figure 3 Final submission score**

The bars show the noise-normalized prediction accuracy of the model (mean of the correlation coefficient) for each region. The colored bars show the prediction accuracy for each subject.

## 4. Conclusions

In this study, we quantitatively investigated how the prediction accuracy of the encoding model depends on the sample size used to build the model and the parameter size of the vision models. We found that the models scaled approximately log-linearly with the sample size used to build the models, and that the tendency was shared across lower to higher visual regions. We also found log-linear scaling for the parameter size of the vision models, especially in the higher visual areas. These results suggest that when building models of the brain's visual pathways using encoding models, further increasing the sample size for model building and the parameter size of the vision models will improve the prediction accuracy.

## Acknowledgements

We were supported by KAKENHI JP18H05522, KAKENHI 23K11167, JST CREST JPMJCR18A5, ERATO JPMJER1801 and MIRAI JPMJMI19D1.